\pdfoutput=1

\documentclass[11pt,twoside,a4paper,cmspaper,final,collab]{cms-tdr}
\def\svnVersion{54335:55781}\def\cmsCernNo{2011-058}\def\cmsCernDate{\today}\def\cmsMessage{Submitted to the ournal of High Energy Physics}

\begin{document}\cmsNoteHeader{SUS-11-002}

\hyphenation{had-ron-i-za-tion}
\hyphenation{cal-or-i-me-ter}
\hyphenation{de-vices}
\RCS$Revision: 55781 $
\RCS$HeadURL: svn+ssh://alverson@svn.cern.ch/reps/tdr2/papers/SUS-11-002/trunk/SUS-11-002.tex $
\RCS$Id: SUS-11-002.tex 55781 2011-05-16 17:05:42Z alverson $
\cmsNoteHeader{SUS-11-002} % This is over-written in the CMS environment: useful as preprint no. for export versions
\title{Search for supersymmetry in events with a lepton, a photon, and large missing transverse energy in $\Pp\Pp$ collisions at $\sqrt{s} = 7\TeV$}% Force line breaks with \\

\address[neu]{Northeastern University}
\address[fnal]{Fermilab}
\address[cern]{CERN}
\author[cern]{The CMS Collaboration}

\date{\today}

\abstract{
   A search is performed for an excess of events, over the standard
   model expectations, with a photon, a lepton, and large missing
   transverse energy in $\Pp\Pp$ collisions at $\sqrt{s} = 7\TeV$. Such
   events are expected in many new physics models, in
   particular a supersymmetric theory that is broken via a
   gauge-mediated mechanism, when the lightest charged and neutral
   gauginos are mass degenerate. The data sample used in this search
   corresponds to an integrated luminosity of 35 pb$^{-1}$ collected
   with the CMS detector at the LHC.  No evidence of such an excess
   above the standard model backgrounds, dominated by $\PW\gamma$
   production, is found. The results are presented as 95\% confidence
   level upper limits on the cross section for a benchmark
   gauge-mediated scenario, and are then converted into exclusion
   limits on the squark, gluino, and wino masses.  }

\hypersetup{%
pdfauthor={CMS Collaboration},%
pdftitle={Search for supersymmetry in events with a lepton, a photon, and large missing transverse energy in pp collisions at sqrt(s) = 7 TeV},%
pdfsubject={CMS},%
pdfkeywords={CMS, physics, software, computing}}

\maketitle %maketitle comes after all the front information has been supplied

\section{Introduction}
\label{sec:intro}
\begin{figure*}[!hbtp]
   \begin{center} \includegraphics[width=0.98\textwidth]{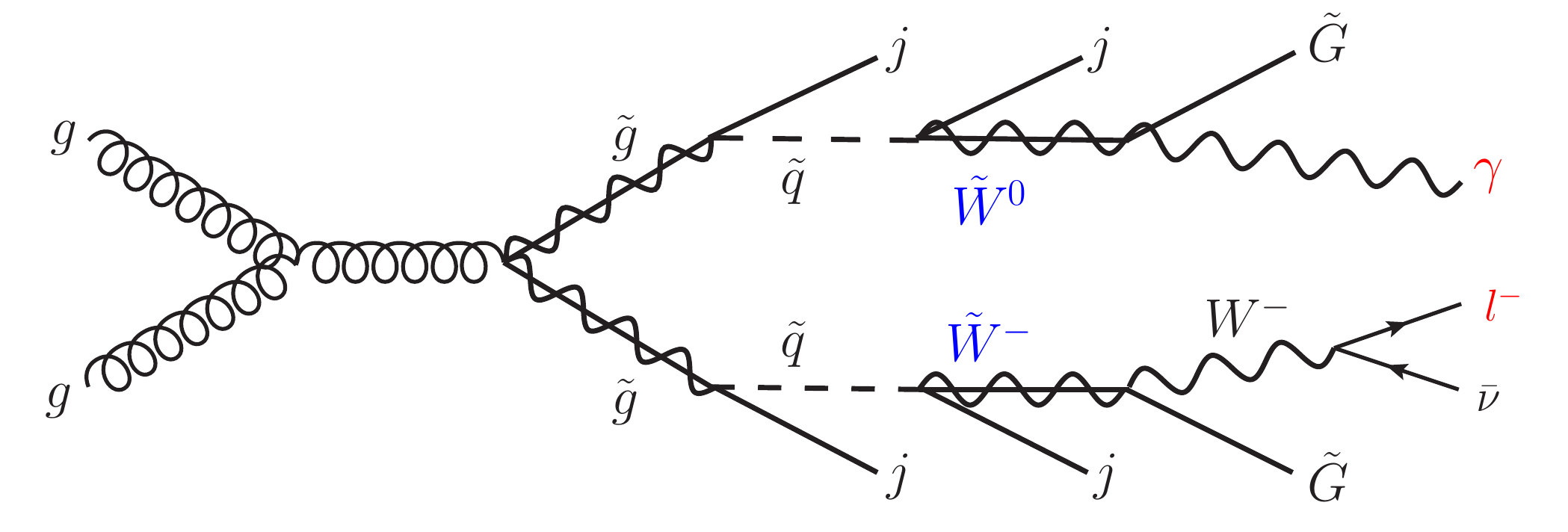}
     \caption{Feynman diagram for an example of a  process with strong production of charged and neutral winos ($\tilde{\PW}^{\pm}$ and $\tilde{\PW}^{0}$) as co-NLSP with multiple jets (j), resulting in the signature $\ell + \gamma + \MET +X$.}
      \label{fig:feyn}
   \end{center}
 \end{figure*}

Supersymmetry (SUSY) is one of the best studied scenarios for physics
beyond the standard model (SM). It alleviates the hierarchy problem
and may provide a path to the grand unification of forces.  Of
particular interest is a scenario with gauge-mediated SUSY breaking
\cite{GGMa,GGMd2,GGMd3,GGMd4,GGMd5,GGMd1,GGMd,Ambrosanio1,Ambrosanio2},
which elegantly addresses the SUSY flavour problem.  In the gauge
mediation scenario, SUSY breaking occurs at energy scales much smaller
than the Planck scale, resulting in the gravitino ($\tilde{G}$) as the
lightest SUSY particle (LSP), which would escape detection and result
in an imbalance of momentum in the plane transverse to the beam
direction (\MET).  In one scenario of general gauge-mediated (GGM)
SUSY~\cite{GGMe,GGMf}, the next-to-lightest SUSY particles (NLSP) are
mostly the superpartners of the $SU(2)_L$ gauge fields (winos).  This
scenario results in a small mass splitting between the charged
($\tilde{\PW}^{\pm}$) and neutral ($\tilde{\PW}^{0}$) winos.  The charged
wino decays promptly into a gravitino and a $\PW$ boson, and the neutral
wino into a gravitino and a photon or a $\Z$ boson ~\cite{GGM-conlsp}.
If R-parity is conserved, coloured SUSY particles are pair-produced,
resulting in the presence of two NLSPs and multiple jets per event. An
example process is shown in Fig.~\ref{fig:feyn}. The event topologies
of interest therefore contain, in addition to \MET, one of the
following signatures: two photons ($\gamma$); lepton(s) ($\ell$) and
one photon; lepton(s) and jets; one photon and jets; or all
jets. Final states with two or more leptons are suppressed because of
the $\PW$ and $\Z$ leptonic branching fractions.

In this paper, we perform the first search for SUSY in the $\ell +
\gamma + \MET + X$ final state produced in $\Pp\Pp$ collisions at
$\sqrt{s} = 7\TeV$ at the Large Hadron Collider (LHC).  The $\ell$
denotes an electron or a muon.  Previous searches for anomalous
production of such events were performed at the Fermilab
Tevatron~\cite{cdfr2,cdfr1_1,cdfr1_2}.

\section{The CMS Detector}
\label{sec:det}
The Compact Muon Solenoid (CMS) is a nearly hermetic, multipurpose
detector and is described in detail elsewhere~\cite{cmsdet}.  Charged
particle trajectories are measured with silicon pixel and strip
tracking detectors encompassed by a 3.8 T superconducting solenoid,
with full azimuthal coverage and $|\eta| < 2.5$. The azimuthal angle
$\phi$ is defined in the plane transverse to the beam direction. The
pseudorapidity $\eta$ is defined as $\eta=-\ln[\tan(\theta/2)]$, where
$\theta$ is the polar angle with respect to the counterclockwise
proton beam direction, with the origin at the center of the detector.
Energy deposits of particles are measured with lead-tungstate crystal
electromagnetic (ECAL) and brass/scintillator ($|\eta| < 3$) or
iron/quartz-fiber ($3 < |\eta| < 5$) hadronic (HCAL) calorimeters.
Muon candidates are identified in gas-ionisation detectors embedded in
the steel return yoke of the magnet.

\section{Candidate Event Selection}
\label{sec:cand}
We use the data sample collected by the CMS detector at the LHC from
March to November 2010 corresponding to an integrated luminosity of 35
pb$^{-1}$~\cite{lumi4}. Events are required to have at least one high
transverse momentum ($\pt$) electron or muon, in addition to a photon.
Events are selected and recorded using a two-tiered trigger system.
We use a suite of single-electron and single-muon triggers with
several thresholds changing to match the rapidly increasing
instantaneous luminosity of the LHC.  The electron candidates are
reconstructed from clusters of energy in the ECAL barrel ($|\eta| <
1.479$) and endcaps ($1.479 < |\eta| < 3$) and are required to match
charged tracks reconstructed in the silicon tracking detectors.  Muons
are reconstructed as charged tracks matched to hit patterns in the
muon detectors.  We further require the leading electron or muon to have
$\pt > 20\GeV$, $|\eta| < 2.1$ and to satisfy a set of identification
and isolation criteria~\cite{wzxs}.  For electrons, the region from
$1.44 < |\eta| < 1.57$ is excluded since it contains services and
cables exiting between the transition of the barrel and endcap
calorimeters and leads to lower quality reconstructed clusters.
Events in which the electron is consistent with having originated from
a photon conversion are rejected.  The efficiency of the triggers for
events with an electron (muon) with $\pt$ above $20\GeV$ is measured
to be $0.98 \pm 0.01$ ($0.968\pm 0.004$).  Lepton
identification and reconstruction efficiencies are measured in data
using $\Z\rightarrow \ell^{+} \ell^{-}$ decays. The ratio of the
lepton efficiencies measured in data to those from Monte Carlo (MC)
simulation (described below) is found to be $0.928 \pm 0.015$ ($0.990
\pm 0.001$) for an electron (muon).  We require the presence of at least
one photon with $\pt > 30\GeV$, $|\eta| < 1.44$, spatially
separated from the lepton by $\Delta R(\ell,\gamma) =
\sqrt{(\Delta\eta)^2 + (\Delta\phi)^2 }>0.4$.  The reconstruction of
photon candidates in the calorimeter is similar to that of the
electrons, with the additional requirement that they do not have
associated hit patterns in the pixel detector consistent with a track.
The data/MC scale factor for the photon efficiency is $0.97 \pm 0.02$.
We demand that the lepton and the photon candidates be isolated from any
other activity in the tracker, ECAL, or HCAL.  The events are required
to contain at least one reconstructed primary vertex within 24 cm of
the geometric center of the detector along the beamline and a
transverse distance with respect to the reconstructed beam axis of less
than 2 cm.  These selections define our candidate sample, which
consists of 264 (126) events in the $\Pe\gamma$ ($\mu\gamma$) channel.  Finally,
in order to minimise the number of expected background events while
retaining high signal efficiency, we select events with $\MET >
100\GeV$.  The \MET is reconstructed using calorimeter energy deposits
and is corrected for the contribution from the muons and for the
expected calorimeter response for each reconstructed hadron
track~\cite{tcmet}.

\section{Background Estimation}
\label{sec:bkg}
The dominant background process is $\PW\gamma$ production, with the
$\PW$ boson decaying leptonically.  This background is estimated using
the {\sc madgraph} MC generator~\cite{Madgraph} with parton showering
using {\sc pythia}~\cite{pythia}.  The result is scaled from the
leading-order (LO) to the next-to-leading-order (NLO) cross section
obtained from the {\sc Baur} NLO $\PW\gamma$ generator~\cite{baurMC}
and using the CTEQ66 NLO parton distribution function (PDF)
sets~\cite{CTEQ66}. For example, for a photon with $\pt = 50\GeV$ the
NLO/LO $K$ factor is 2.3.  The NLO calculation is in good agreement
with the $\PW\gamma$ production cross section measured by the CMS
collaboration~\cite{ewk_wgamma}, over the relevant phase space of that
measurement, giving confidence in the NLO prediction used here.  The
MC events are processed with the full CMS detector simulation based on
the Geant4~\cite{geant4} package. To estimate the uncertainty on
the $K$ factor due to higher-order corrections, we vary the
factorisation and renormalisation scales by a factor of two and
observe a variation in the cross section of approximately $20$\%.
Additional systematic uncertainties on this background estimate
include the uncertainties on the trigger efficiencies and data/MC
scale factors described above, the PDF set estimated using the PDF4LHC
working group procedure (2\%)~\cite{pdf4lhc}, and the integrated
luminosity measurement (4\%)~\cite{lumi4}.

The next most important instrumental background originates from SM
processes that have misidentified leptons and photons.  Data-driven
methods are used to determine all backgrounds involving object
misidentification~\cite{sus_gg}. We consider sources of background in
which a jet or an electron is misidentified as a photon (jet $\rightarrow
\gamma$ and $\Pe \rightarrow \gamma$) or in which a jet is
misidentified as a lepton.  The dominant SM processes that contribute
to the jet $\rightarrow \gamma$ misidentification background include $\PW+$jets and
QCD multijet production.  The jet $\rightarrow \gamma$ background is
determined by selecting a sample of events that satisfy the criteria
of our $\ell \gamma$ candidate sample except that the photon is
required to fail either the isolation or electromagnetic shower shape
criterion. We then weight this sample by the probability of
misidentifying a jet as a photon~\cite{led_gg} to estimate the total contribution of
this background to the signal sample.  The SM processes
contributing to the $\Pe \rightarrow \gamma$ misidentification background are
predominantly $\Z$ and $t \bar{t}$ production.  The $\Pe \rightarrow
\gamma$ background is determined in a similar way to the jet
$\rightarrow \gamma$ background, by selecting a sample of events that
satisfy the criteria of our $\ell
\gamma$ candidate sample but with the additional requirement that the
photon be matched to a pattern of hits in the pixel tracker. We then
weight this sample by the probability of misidentifying an electron as
a photon ($f_{\gamma_\Pe}$) to estimate the total contribution of this
background to the signal sample. The probability, $f_{\gamma_\Pe}$, is
determined from $\Z\rightarrow \Pe^{+} \Pe^{-}$ data to be $0.014 \pm
0.004$.

The strategy for determining QCD contributions which have poorly
measured \MET is to select events with kinematic properties similar to
the candidate events, but which are known to have no genuine \MET.  A
control sample of events dominated by $\Z\rightarrow \Pe^{+}\Pe^{-}$
decays is used.  These events are further reweighted to reproduce the
$\ell\gamma$ sample kinematics.  In particular, we reproduce the
$\ell\gamma$ transverse energy distribution and, therefore, the
transverse energy of the hadronic recoil against the $\ell\gamma$
candidates, ensuring a correct description of the \MET. In order to
describe the distribution of the transverse mass of the lepton and
\MET system, defined as $M_T = \sqrt{2 \cdot p_{T_{\ell}} \cdot \MET
\cdot (1 - \cos (\phi_{\ell} - \phi_{\MET}))}$, the lepton $p_T$
distribution is also reweighted.  We determine the background
contribution by normalising the control sample in such a way that the
total background matches the number of selected events in data with
\MET less than $30\GeV$. Tables~\ref{tab_e3_met} and
\ref{tab_mu3_met} summarise the breakdown of the backgrounds for $\MET
> 40\GeV$ and for our optimal selection of $\MET > 100\GeV$. The
dominant sources of systematic uncertainties are due to the limited
precision of the data-driven methods and the uncertainties on the
corresponding misidentification rate measurements.

\begin{table*}[htbp]
\begin{center}
\caption{Event counts for data and expected background in the $\Pe\gamma$ channel. The uncertainty on the total background includes
   the anti-correlation between the uncertainties on the individual
   components due to the scaling of the QCD background to the events with $\MET < 30\GeV$ as described in the text.  The expected event yields from  the SUSY benchmark point GMC (Section~\ref{sec:res}) are also shown. \label{tab_e3_met}}
\begin{tabular}{|c|c|c|c|}
\hline \hline
 & No \MET selection & $\MET > 40\GeV$ & $\MET > 100\GeV$ \\
\hline \hline
$\PW\gamma$ & $44.5 \pm 9.2$     & $16.1    \pm 3.4  $  & $1.68    \pm 0.42 $ \\
jet$\rightarrow \gamma$ & $20.3 \pm 4.5$     & $3.1     \pm 0.9  $  & $0.02    \pm 0.02 $ \\
$\Pe\rightarrow \gamma$ & $70.5 \pm  19.1$     & $0.3     \pm 0.1  $  & $0.04    \pm 0.03 $ \\
QCD                   & $134 \pm 28$    & $0.4     \pm 0.2  $  & $0.00    \pm 0.00 $ \\
\hline
Total background & $269 \pm 18$    & $19.9    \pm 3.7  $  & $1.74    \pm 0.43 $ \\
\hline
data & 264 & 16 & 1 \\
\hline \hline
SUSY GMC prediction  & 3.94 $\pm$ 0.79 & 3.76 $\pm$ 0.75 & 2.79 $\pm$ 0.56 \\
\hline \hline
\end{tabular}
\end{center}
\end{table*}

\begin{table*}[htbp]
\begin{center}
\caption{Event counts for data and expected background in the $\mu\gamma$ channel.  The uncertainty on the total background includes
   the anti-correlation between the uncertainties on the individual
   components due to the scaling of the QCD background to the events with $\MET  < 30\GeV$ as described in the text. The expected event yields from the SUSY benchmark point GMC (Section~\ref{sec:res}) are also shown. \label{tab_mu3_met}}
\begin{tabular}{|c|c|c|c|}
\hline \hline
 & No \MET selection & $\MET > 40\GeV$ & $\MET > 100\GeV$ \\
\hline \hline
$\PW\gamma$ & $44.8 \pm 9.3$     & $15.9    \pm 3.4  $  & $1.40    \pm 0.37 $ \\
jet$\rightarrow \gamma$ & $18.0 \pm 4.0$     & $3.7     \pm 1.1  $  & $0.10    \pm 0.09 $ \\
$\Pe\rightarrow \gamma$ & $1.2 \pm 0.4$      & $0.6     \pm 0.2  $  & $0.09    \pm 0.04 $ \\
QCD                    & $58.3 \pm 15.1$     & $0.2     \pm 0.1  $  & $0.00    \pm 0.00 $ \\
\hline
Total background & $122.3 \pm 12.3$    & $20.4    \pm 3.7  $  & $1.59    \pm 0.39 $ \\
\hline
data & 126 & 27 & 1 \\
\hline \hline
SUSY GMC prediction & 5.12 $\pm$ 1.02 & 4.84 $\pm$ 0.96 & 3.66 $\pm$ 0.73 \\
\hline \hline
\end{tabular}
\end{center}
\end{table*}

\section{Results}
\label{sec:res}

\begin{figure*}[!hbtp]
  \begin{center} \includegraphics[width=0.98\textwidth]{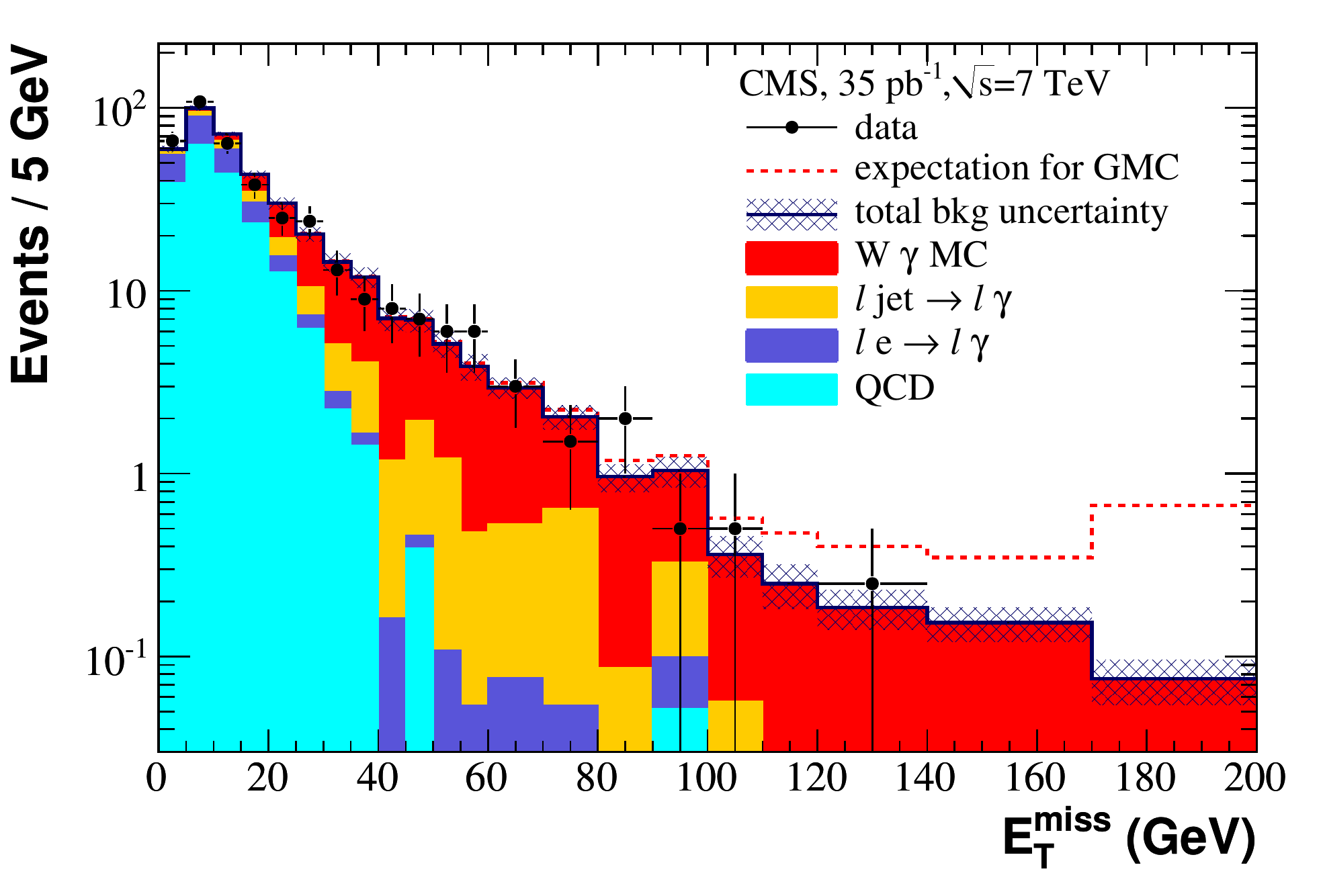}
    \includegraphics[width=0.98\textwidth]{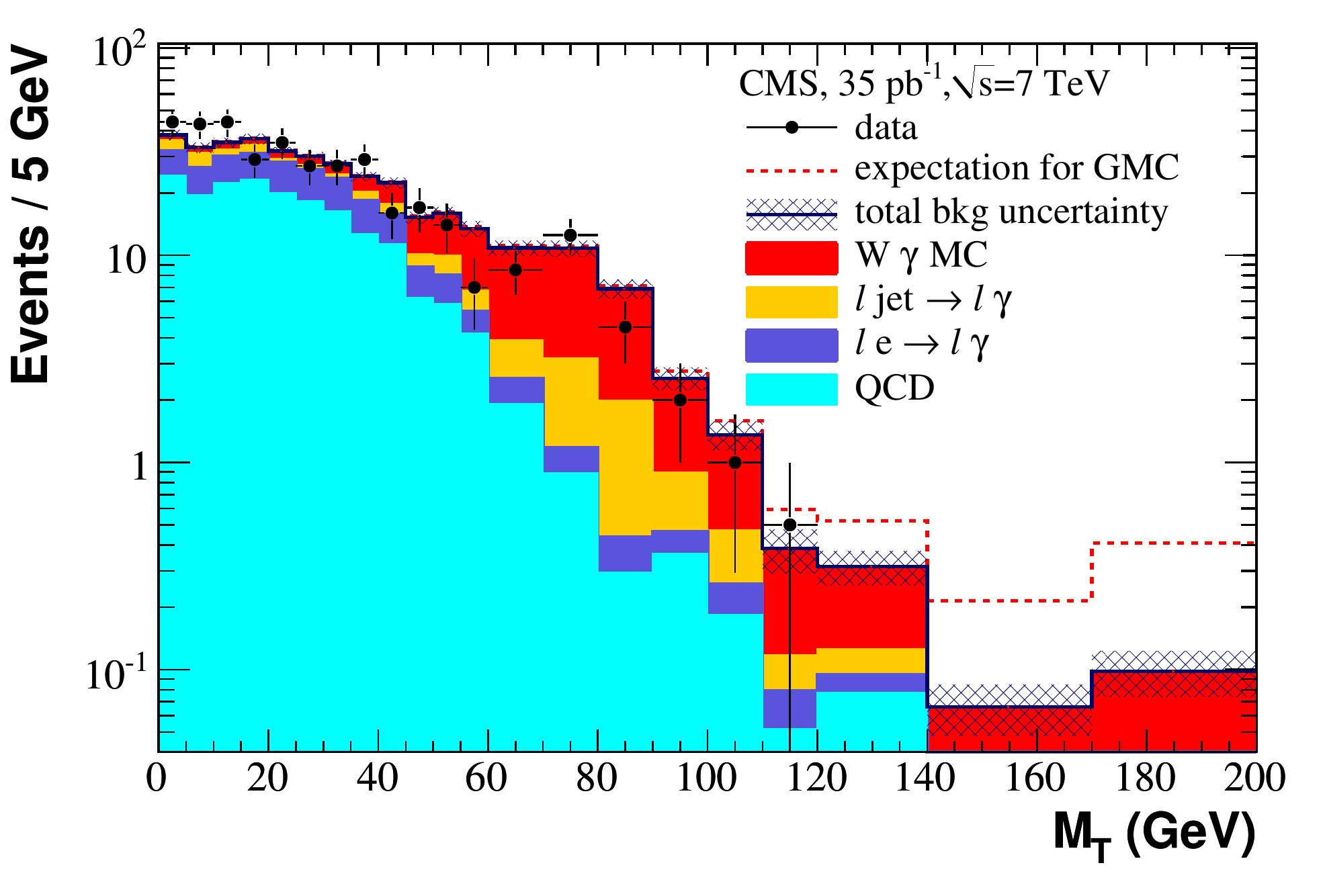} \caption{The \MET
    (top) and $M_T$ distributions (bottom) for the combined $\Pe\gamma$
    and $\mu\gamma$ samples, with logarithmic scales on the vertical axes;
    the last bin shown in both distributions includes the overflow
    events.  The black points represent the 35 pb$^{-1}$ data sample,
    and the coloured histograms show the individual background
    components.  The total background and its uncertainty are
    represented by the shaded band.  The
    expectation from the SUSY benchmark point GMC is shown by the red
    dashed line.}  \label{fig:met_mt} \end{center}
\end{figure*}

The \MET distribution is shown in Fig.~\ref{fig:met_mt} (top) for the
$\Pe\gamma$ and $\mu\gamma$ channels combined, together with the sum
of the expected background contributions, as well as the expectation
of the signal from a SUSY benchmark point (referred to as GMC,
described below) for comparison.  Figure~\ref{fig:met_mt} (bottom)
shows the $M_T$ distribution.  The data agree well with the SM
expectations and do not exhibit any evidence for an excess of events
with high \MET.

We performed several cross-checks to support our modeling of
the SM backgrounds.  For example, events with $M_T$ above $70\GeV$
have small QCD background, as shown in Fig.~\ref{fig:met_mt} (bottom),
and we verified that we can describe all the kinematic distributions
in these events with the sum of the expected backgrounds, except the contribution
from QCD.  We also checked that the photon candidates have a
timing consistent with having originated from the primary vertex,
to verify that non-collision backgrounds are negligible.  Finally, we
verified that our QCD background determination procedure works in the
presence of a large range of SUSY signals by mixing simulated SUSY
events into the data sample and recalculating the QCD background; the
change in its value was found to be negligible.

Since we observe no evidence of an excess of events with high \MET
compared to the SM expectations, we interpret our results by placing
95\% confidence level (CL) lower limits on GGM SUSY particle masses in
a benchmark model.  The GGM SUSY~\cite{GGM-conlsp} signal is simulated
using the {\sc pythia}~\cite{pythia} MC generator version 6.420 with
tune D6T~\cite{d6t}, and assuming that the squark mass
($m(\tilde{q})$) is approximately equal to the gluino mass
($m(\tilde{g})$), with the charged and neutral winos as co-NLSPs, and
$\tan{\beta} = 2$.  This search focuses on the $m(\tilde{W}^{0}) >
100\GeV$ region, which is beyond the sensitivity of the previous
collider experiments~\cite{cdfr2,cdfr1_1,cdfr1_2,shih}.  We therefore
generate points with $m(\tilde{g})$ and $m(\tilde{q})$ in the range of
350--$800\GeV$ and $m(\tilde{\PW}^{0}) \approx m(\tilde{\PW}^{\pm})$ in the range of
100--$480\GeV$.  The generated events are processed with the
full CMS detector simulation using the Geant4 package.  The NLO
$K$ factors were calculated using {\sc prospino}~\cite{prospino} and
are found on average to be 1.4 with a $\sim 20$\% variation.  As an
example of a SUSY benchmark model, we use the point generated with
$m(\tilde{g}) = m(\tilde{q}) = 450\GeV$ and $m(\tilde{\PW}^{0})
\approx m(\tilde{\PW}^{\pm}) = 195\GeV$, referred to as GMC.

\begin{figure*}[!htbp]
  \begin{center}
    \includegraphics[width=0.98\textwidth]{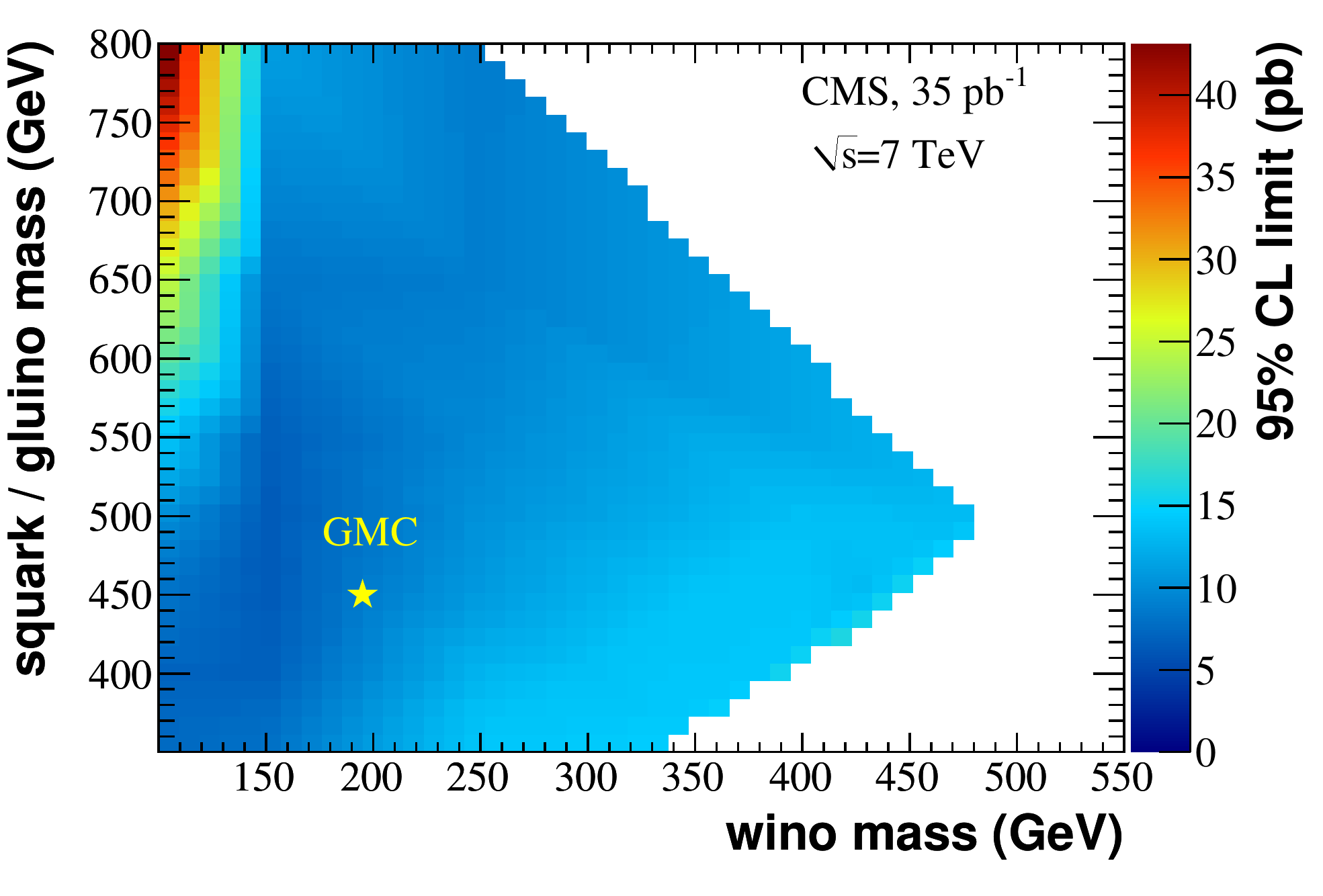}
    \caption{The observed 95\% CL upper limits on the cross section as a function of the squark~/~gluino mass (since $m(\tilde {q})\approx m(\tilde {g})$) and the wino mass is shown for the range of MC points generated for this analysis.   The star indicates the SUSY benchmark point GMC.}
    \label{fig:xseclim}
  \end{center}
\end{figure*}

\begin{figure*}[!htbp]
  \begin{center}
    \includegraphics[width=0.98\textwidth]{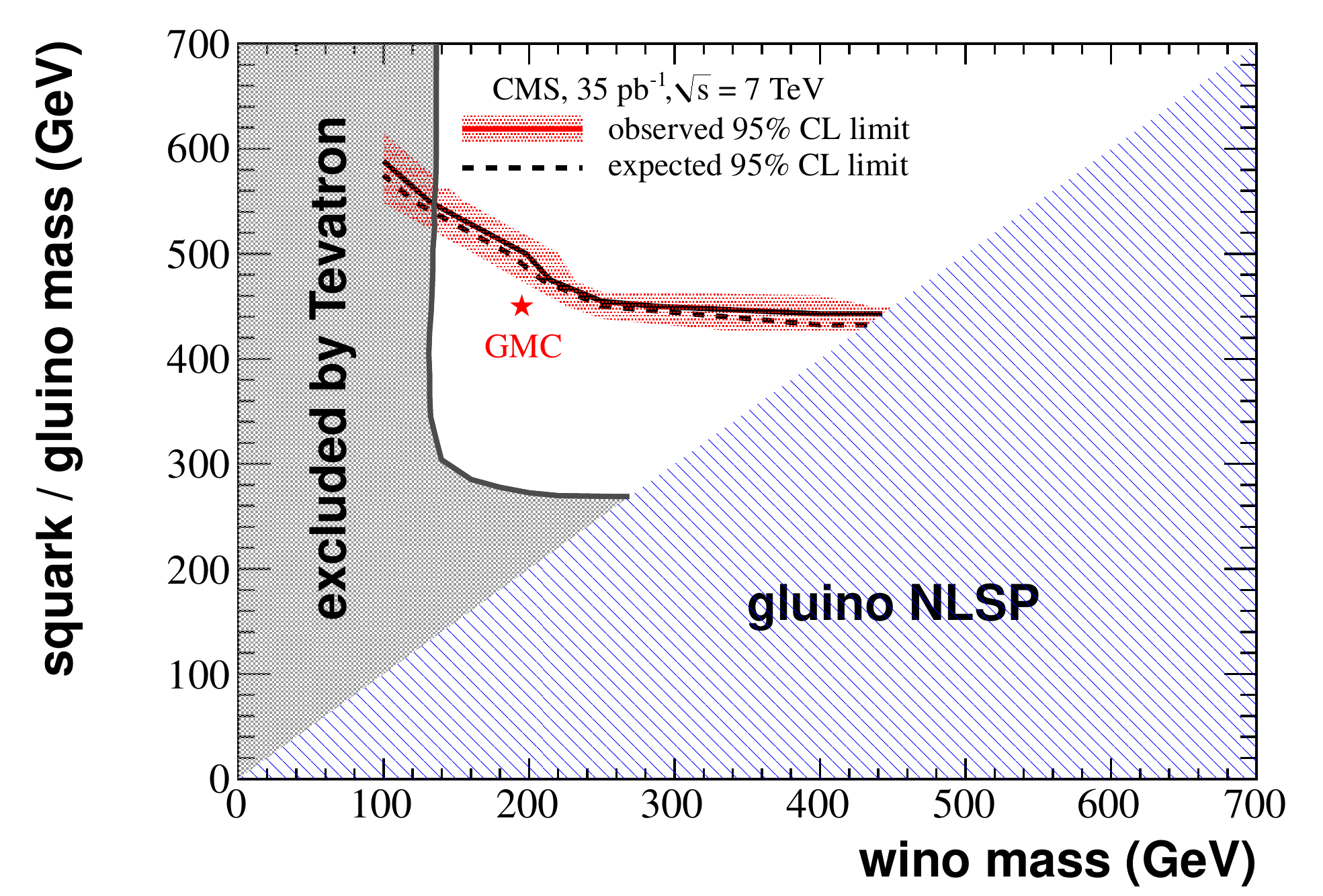}
    \caption{The 95\% CL exclusion region as a function of the
    squark~/~gluino mass (since $m(\tilde {q})\approx m(\tilde
    {g})$) and the wino mass.  The area below the curve is excluded by
    this analysis.  The shaded region around the central value of the
    exclusion contour reflects the uncertainty in the NLO predicted
    cross section.  The dashed line represents the expected 95\% CL
    limit from the MC simulation.  The region excluded by the Tevatron
    in this channel~\cite{cdfr2,cdfr1_1,cdfr1_2} as interpreted
    in~\cite{shih} is also shown.  The star indicates the SUSY
    benchmark point GMC, which lies in the excluded region.  The region
    for the scenario of the gluino as the NLSP is specified by
    the hatched area, which is not accessible to this analysis.}
    \label{fig:massexcl} \end{center}
\end{figure*}

We use a Bayesian approach to determine the 95\% CL upper limits on
the cross sections in the benchmark SUSY model described above,
combining the likelihoods of the $\Pe\gamma$ and $\mu\gamma$ channels.
The inputs for both channels are the sum of the data-driven
backgrounds, the $\PW\gamma$ background as determined from the MC
simulation and corrected by the appropriate data/MC scale factors,
and, for each MC SUSY point, the efficiency times acceptance,
corrected for the appropriate data/MC scale factors.  Each of these
inputs for a given channel has log-normal distributed nuisance
parameters that vary by the uncertainty in each quantity.  We consider
the 4\% uncertainty on the luminosity as an additional nuisance
parameter.  As an example of the calculation, the efficiency times
acceptance times branching ratio for the GMC benchmark point for
$\Pe\gamma$ ($\mu\gamma$) is $0.0063 \pm 0.0003$ ($0.0083 \pm
0.0004$).  The 95\% CL upper limit on the cross section obtained for
this point is 8.7 pb.  The limit is less than the predicted NLO cross
section for this point, (12.7 $\pm$ 2.6) pb, with uncertainties due
to the PDF~\cite{pdf4lhc} (10\%) and the renormalization scale
(18\%). We therefore exclude this point with more than 95\% confidence.
The observed 95\% CL upper limits on the cross section are shown in
Fig.~\ref{fig:xseclim} as a function of the squark~/~gluino and the
wino masses for the kinematic selections applied in this analysis
(an electron with $\pt > 20\GeV$, $|\eta| < 1.44$ or $1.57 < |\eta| <
2.1$; a muon with $\pt > 20\GeV$ and $|\eta| < 2.1$; a photon with $\pt
> 30\GeV$, $|\eta| < 1.44$; events with $\MET > 100\GeV$).
Figure~\ref{fig:massexcl} displays the resulting 95\% CL exclusion
region in the squark~/~gluino and wino mass plane, where masses in the
area below the curve are excluded.  The shaded region around the
central value of the exclusion contour represents the variation of the
predicted cross section by one standard deviation.  Also shown is the
expected 95\% CL exclusion as well as the the region excluded by
previous searches at the Tevatron in the same
channel~\cite{cdfr2,cdfr1_1,cdfr1_2}, as interpreted in~\cite{shih}.

\section{Conclusions}
\label{sec:concl}

We performed the first search in proton-proton collisions at $\sqrt{s}
= 7\TeV$ for an excess of events with a photon, a lepton, and large
\MET over the SM expectation.  Using a data sample corresponding to an
integrated luminosity of 35 pb$^{-1}$, collected with the CMS detector
at the LHC, we find no evidence of such an excess.  Our results are
interpreted as 95\% CL lower limits on GGM SUSY particle masses in a
benchmark model.  Additionally, the 95\% CL upper limits on the cross
sections of generic SUSY production as a function of squark~/~gluino
and wino masses are also provided, allowing other SUSY scenarios to be
constrained.  These limits are the most stringent to date for a large
range of parameters in this GGM SUSY scenario.

\section*{Acknowledgements}

We thank David Shih and Joshua Ruderman for stimulating
discussions, support, and help with the production of the SUSY Monte Carlo
samples.  We wish to congratulate our colleagues in
the CERN accelerator departments for the excellent performance of the
LHC machine.  We thank the technical and administrative staff at CERN
and other CMS institutes, and acknowledge support from: FMSR
(Austria); FNRS and FWO (Belgium); CNPq, CAPES, FAPERJ, and FAPESP
(Brazil); MES (Bulgaria); CERN; CAS, MoST, and NSFC (China);
COLCIENCIAS (Colombia); MSES (Croatia); RPF (Cyprus); Academy of
Sciences and NICPB (Estonia); Academy of Finland, ME, and HIP
(Finland); CEA and CNRS/IN2P3 (France); BMBF, DFG, and HGF (Germany);
GSRT (Greece); OTKA and NKTH (Hungary); DAE and DST (India); IPM
(Iran); SFI (Ireland); INFN (Italy); NRF and WCU (Korea); LAS
(Lithuania); CINVESTAV, CONACYT, SEP, and UASLP-FAI (Mexico); PAEC
(Pakistan); SCSR (Poland); FCT (Portugal); JINR (Armenia, Belarus,
Georgia, Ukraine, Uzbekistan); MST and MAE (Russia); MSTD (Serbia);
MICINN and CPAN (Spain); Swiss Funding Agencies (Switzerland); NSC
(Taipei); TUBITAK and TAEK (Turkey); STFC (United Kingdom); DOE and
NSF (USA).

\bibliography{auto_generated}   % will be created by the tdr script.

\cleardoublepage\appendix\section{The CMS Collaboration \label{app:collab}}\begin{sloppypar}\hyphenpenalty=5000\widowpenalty=500\clubpenalty=5000\textbf{Yerevan Physics Institute,  Yerevan,  Armenia}\\*[0pt]
S.~Chatrchyan, V.~Khachatryan, A.M.~Sirunyan, A.~Tumasyan
\vskip\cmsinstskip
\textbf{Institut f\"{u}r Hochenergiephysik der OeAW,  Wien,  Austria}\\*[0pt]
W.~Adam, T.~Bergauer, M.~Dragicevic, J.~Er\"{o}, C.~Fabjan, M.~Friedl, R.~Fr\"{u}hwirth, V.M.~Ghete, J.~Hammer\cmsAuthorMark{1}, S.~H\"{a}nsel, M.~Hoch, N.~H\"{o}rmann, J.~Hrubec, M.~Jeitler, W.~Kiesenhofer, M.~Krammer, D.~Liko, I.~Mikulec, M.~Pernicka, H.~Rohringer, R.~Sch\"{o}fbeck, J.~Strauss, A.~Taurok, F.~Teischinger, P.~Wagner, W.~Waltenberger, G.~Walzel, E.~Widl, C.-E.~Wulz
% ...
\vskip\cmsinstskip
% ...
1:~~Also at CERN, European Organization for Nuclear Research, Geneva, Switzerland\\
% ...

\end{sloppypar}
\end{document}